\documentclass[preprint]{epl}

\newcommand{\bra}{\left\langle}
\newcommand{\ket}{\right\rangle}

\newcommand{\pder}[2]{\frac{\partial #1}{\partial  #2}}

\newcommand{\e}{{\rm e}}

\newcommand{\Zd}{Z_{\rm d}}

\newcommand{\hks}{h_{\rm ks}}

\title{Computation of the Kolmogorov-Sinai entropy using statistitical
mechanics: Application of an exchange Monte Carlo method}
\shorttitle{Computation of the KS entropy}

\author{Shin-ichi Sasa\inst{1}
\thanks{E-mail:\email{sasa@jiro.c.u-tokyo.ac.jp}} 
\and Kumiko Hayashi\inst{1}
\thanks{E-mail:\email{hayashi@jiro.c.u-tokyo.ac.jp}} }
\institute{
\inst{1}
Department of Pure and Applied Sciences, University of Tokyo, Komaba, Tokyo 153-8902, Japan}


\begin{document}

\maketitle

\begin{abstract}
We propose a method for computing the Kolmogorov-Sinai (KS) entropy 
of chaotic systems. In this method,  the KS entropy is expressed as 
a statistical average over the canonical ensemble for a Hamiltonian
with many ground states. This Hamiltonian is constructed
directly from an evolution equation that exhibits chaotic dynamics. 
As an example, we compute the KS entropy for a chaotic repeller
by evaluating the thermodynamic entropy of a system with many 
ground states. 
\end{abstract}

%%%%%%%%%%%%%%%%%%%%%%
%% introduction     %%
%%%%%%%%%%%%%%%%%%%%%%

%%\baselineskip=20pt

%%  motivation  %%

Chaotic systems are characterized by properties of $O(\e^{N}) $
distinguishable trajectories over a time  interval $N$ 
in a phase space in which the limit of observational accuracy is 
given by some finite value $\epsilon$\cite{chaos1,chaos2,Dorfman}.   
In particular, if these trajectories are coded as sequences of symbols, 
the optimal compression ratio of such sequences is an important quantity 
from the point of view of information theory \cite{BS,CT}. This quantity 
corresponds to  the Kolmogorov-Sinai (KS) entropy, $\hks$.

Although  the KS entropy provides a simple characterization of
chaotic systems, its direct numerical evaluation is not simple.
One reason for this is that a precise mathematical definition 
of distinguishable trajectories 
is difficult to investigate numerically \cite{chaos1}.  
However, 
this difficulty can be overcome when  a set of  periodic orbits 
alone can provide information that is sufficient to obtain a 
useful characterization of the set of all distinguishable trajectories  
\cite{Tomita,GP,Ci}.  However, even in such cases, the difficulty 
involved 
in numerical computations remains, because the number of periodic orbits of 
period $N$  increases exponentially as a function of $N$. 

Here, let us recall that in the statistical mechanics of glassy 
systems \cite{glass} and  decision problems \cite{opt}, the number of 
explored states increases exponentially as a function of the system 
size.  Recently,  efficient methods for computing  thermodynamic quantities 
of glassy systems have been developed \cite{berg,exmc} and 
applied to several problems \cite{magic,number,queen}. 
Inspired by this success, in this paper, we propose  a  
method for computing the KS entropy employing a quantity that 
can be calculated in a system with many ground states. 

%% specific introduction 

Specifically, we consider a map $\Gamma_{n+1}=F(\Gamma_n)$, 
with $\{\Gamma_n\} 
\in \Omega \subset {\mathbf R}^d$, and assume that its 
trajectories can be 
coded with sequences of $D$ symbols. The frequency distribution of 
an $N$ symbol 
sequence $[s^k]=(s_1^k,s_2^k,\cdots,s_N^k)$, with the natural measure 
for the initial conditions, is denoted by $\mu([s^k])$, where $s_n^k \in  
\{1,\cdots , D\}$, and $k$ is the index identifying 
 the sequence. Letting $M(N)$
be the total number of possible symbol sequences, the KS entropy, 
$\hks$, is defined in terms of  the Shannon entropy rate \cite{BS} as  
\begin{equation}
\hks\equiv -\lim_{N \to \infty} \frac{1}{N} \sum_{k=1}^{M(N)} 
\mu([s^k]) \log \mu([s^k]).
\label{hks:def}
\end{equation}
Note that $\hks/\log(D)$ represents the optimal compression 
ratio of symbol sequences \cite{CT}. 

%% KS entropy (periodic)

In order to study simple cases, we further assume that 
for each $[s^k]$ there exists an $N$-periodic orbit 
$\Gamma_1^k,\Gamma_2^k,\cdots,\Gamma_N^k,\Gamma_1^k$   
such that the symbol corresponding to $\Gamma_i^k$ is $s_i^k$ 
for all $i \le N$. 
Then there is a one-to-one correspondence between the set of all 
symbol sequences $\{[s^k]\}$ and the set of all $N$-periodic orbits 
$\{[\Gamma^k]\}$.  
Furthermore, when $N$ is sufficiently large, $\mu([s^k])$ is 
accurately approximated by the probability
\begin{equation}
p_k \equiv \frac{1}{\Zd(N)}\e^{-N \lambda([\Gamma^k ])},
\label{pk:def}
\end{equation}
where $\lambda([\Gamma^k])$ is the volume expansion ratio in the unstable
directions around the periodic orbit $[\Gamma^k]$. 
The normalization constant $\Zd$ is determined by the condition 
$\sum_{k=1}^{M(N)}p_k=1$.  
In terms of  $p_k$,  the KS entropy is expressed as
\begin{equation}
\hks=-\lim_{N \to \infty} \frac{1}{N}
\sum_{k=1}^{M(N)} p_k \log p_k.
\label{hks:def2}
\end{equation}
Note that the relation $\lim_{N \to \infty} \log \Zd(N)/N =0$ holds 
for  many 
physically interesting systems. In this case, the expression 
(\ref{hks:def2}) leads to the  useful relation  (Pesin's formula) 
that  $\hks$ is equal to the sum of the positive  Lyapunov exponents 
\cite{chaos1,Dorfman}. 
However, there are systems to which this formula cannot be applied.
A typical example is  a chaotic repeller. Because the KS entropy of
a repeller is related to  transport properties of nonequilibrium 
phenomena described by deterministic models \cite{GN}, 
it is important to study a computational method for $\hks$ 
on the basis of (\ref{hks:def2})  that does not employ Pesin's formula.

%%%%%%%%%%%%%%%%%%%%%%%%%%%%%%
% basic formulation          %
%%%%%%%%%%%%%%%%%%%%%%%%%%%%%%

\section{Basic formula:}

Here and hereafter, we use 
$[\Gamma]=(\Gamma_1,\cdots,\Gamma_N)$ to represent 
an array of $N$ phase space points, not a trajectory  given by the 
map $\Gamma_{n+1}=F(\Gamma_n)$.  We also introduce $\Gamma_0$ as 
$\Gamma_0=\Gamma_N$ for later convenience.  
Let $\Lambda_m([\Gamma])$, with $1 \le m \le d$, 
be the eigenvalues of the matrix 
$F'(\Gamma_N)F'(\Gamma_{N-1})\cdots F'(\Gamma_1)$,
where $F'(\Gamma_n)$ is the $d \times d$ Jacobi matrix 
determined from the  map $\Gamma_{n+1}=F(\Gamma_n)$. 
The volume expansion ratio $\lambda([\Gamma^k])$ in the unstable 
directions around the periodic orbit $[\Gamma^k]$ is defined as 
\begin{equation}
\lambda([\Gamma^k])\equiv \frac{1}{N} \sum_{m;|\Lambda_m([\Gamma^k])| > 1 }
\log |\Lambda_m([\Gamma^k])|.
\label{lamdef}
\end{equation}
Here, we assume that all periodic orbits 
are hyperbolic  and hence satisfy  
$|\Lambda_m |\not =1$ for  $1 \le m \le d$.  The volume expansion ratio 
$\lambda([\Gamma])$ for an arbitrary array $[\Gamma]$ can be 
defined  by replacing $\Gamma^k$  in (\ref{lamdef}) with $\Gamma$.

For  an arbitrary function  $A([\Gamma])$, we define the quantity
\begin{equation}
I(A)\equiv  \sum_{k=1}^{M(N)} A([\Gamma^k]) e^{-N \lambda([\Gamma^k])}.
\label{I:def}
\end{equation}
Substituting  (\ref{pk:def}) into (\ref{hks:def2}) and using (\ref{I:def}),
we rewrite $\hks$ as
\begin{equation}
\hks= \lim_{N \to \infty} 
\left[ \frac{I(\lambda)}{I(1)}+\frac{1}{N} \log I(1) \right].
\label{hks:def3}
\end{equation}
We now seek to express  $I(A)$ in a form that can be computed
easily. We first rewrite (\ref{I:def}) as 
\begin{equation}
I(A)=\sum_{k=1}^{M(N)}
\left[\prod_{n=1}^N \int d\Gamma_n \delta(\Gamma_n-\Gamma_n^k)  \right] 
A([\Gamma]) e^{-N \lambda([\Gamma])}.
\label{I:def2}
\end{equation}
Then, defining $\phi_n([\Gamma_n])\equiv \Gamma_n-F(\Gamma_{n-1})$,
$n=1,2, \cdots, N$, we obtain the formula 
\begin{equation}
\prod_{n=1}^N \delta(\phi_n)=\sum_{k=1}^{M(N)}
\frac{\prod_{n=1}^N \delta(\Gamma_n-\Gamma_n^{k})}{|J([\Gamma])|},
\label{delta}
\end{equation}
where 
\begin{equation}
J([\Gamma])=\det 
\frac{\partial(\phi_1,\cdots,\phi_N)}{\partial(\Gamma_1,\cdots,\Gamma_N)}.
\end{equation}
Here, we note that $ J([\Gamma^k])\not=0$ for all hyperbolic 
periodic orbits, because in this case it can be proved that 
\begin{equation}
J([\Gamma]) = \prod_{m=1}^d \left(1-\Lambda_m([\Gamma]) \right).
\label{matform}
\end{equation}
Also, from the expression (\ref{matform}), we can derive 
\begin{equation}
|J([\Gamma^k])| =\e^{N\lambda([\Gamma^k])+o(N)}.
\label{Jexp}
\end{equation}
Then, expressing  the delta function as 
\begin{equation}
\delta(x)=\lim_{\beta \to \infty}
\sqrt{\frac{\beta}{\pi}}\e^{-\beta x^2},
\end{equation}
and ignoring the sub-exponential contribution, 
we can rewrite $I(A)$ as 
\begin{equation}
I(A)=\lim_{\beta \to \infty} \left( \frac{\beta}{\pi} \right)^{Nd/2}
\int d\Gamma_1\cdots d\Gamma_N
\e^{-\beta H([\Gamma])}A([\Gamma]),
\label{I:def4}
\end{equation}
with 
\begin{equation}
H([\Gamma]) = \sum_{n=1}^N |\Gamma_n-F(\Gamma_{n-1})|^2.
\label{hamil:def}
\end{equation}

The expression (\ref{I:def4}) leads us to introduce a canonical ensemble 
in the form 
\begin{equation}
p_{\rm c}([\Gamma])=\frac{1}{Z(\beta)}\e^{-\beta H([\Gamma])}, 
\label{can}
\end{equation}
with the partition function
\begin{equation}
Z(\beta)=\int d\Gamma_1\cdots d\Gamma_N
\e^{- \beta H([\Gamma])} .
\label{part}
\end{equation}
Then, (\ref{I:def4}) becomes
\begin{equation}
I(A)
=\lim_{\beta \to \infty} \left( \frac{\beta}{\pi} \right)^{Nd/2}
\bra A \ket_\beta Z(\beta),
\label{I:def5}
\end{equation}
where $\bra \ \ket_\beta$ represents the average taken with respect 
to (\ref{can}). 
Substituting this result into (\ref{hks:def3}), we
obtain the formula
\begin{equation}
\hks= \lim_{N \to \infty} \lim_{\beta \to \infty}
\left[\bra \lambda \ket_\beta+ \frac{d}{2}
\log \left(\frac{\beta}{\pi} \right)+\frac{1}{N}\log Z(\beta)
\right].
\label{hks:def4}
\end{equation}
Recall that the Hamiltonian (\ref{hamil:def}) is non-negative and is 
equal to  
zero when $[\Gamma]$ coincides with a periodic orbit $[\Gamma^k]$.
Also, the number of periodic orbits $M(N)$ is of order $\e^{N}$
for a chaotic map $F$. These facts imply that the KS entropy, $\hks$, 
can be obtained using the formalism of statistical mechanics with  
the Hamiltonian (\ref{hamil:def}),  which possesses  
$O(\e^{N})$ ground states. 

%%%%%%%%%%%%%%%%%%%%%%%%%%%%%%
% Computational  method        %
%%%%%%%%%%%%%%%%%%%%%%%%%%%%%%

\section{Computational method:}

Let us define the entropy function $S(E)$ through the relation 
\begin{equation}
\e^{S(E)}dE = \int d\Gamma_1 \cdots d \Gamma_N 
{}_{E \le H([\Gamma]) \le E+dE}.
\end{equation}
We then evaluate the partition function $Z(\beta)$ as
\begin{equation}
Z(\beta) =  \e^{S(E_*(\beta))-\beta E_*(\beta)+O(\log (N))}, 
\label{zest}
\end{equation}
where $E_*(\beta)$ is defined as
\begin{equation}
\pder{S(E_*)}{E_*}=\beta.
\end{equation}
Then, from a standard argument employed in  statistical mechanics, 
the relation $E_*(\beta)=\bra H \ket_\beta$ is obtained, and 
the quantity $\tilde S(\beta)\equiv S(E_*(\beta))$ is calculated 
from the identity 
\begin{equation}
-\beta \pder{\tilde S(\beta)}{\beta} = C(\beta), 
\label{sc}
\end{equation}
where $C(\beta)$ represents the heat capacity defined by
\begin{equation}
C(\beta)\equiv \beta^2 \bra (\delta E)^2 \ket_\beta.
\end{equation}

Here, because $C_\infty\equiv \lim_{\beta \to \infty} C(\beta)$ 
takes a nonzero value, $\tilde S(\beta)$ diverges as 
 $\sim \log \beta$ in the limit  $\beta \to \infty$. 
Therefore, 
in order to avoid a singularity, we define $\tilde S_0(\beta)$ 
according to 
\begin{equation}
\tilde S(\beta)=-C_\infty \theta(\beta-1)\log\beta +\tilde S_0(\beta),
\label{s0def}
\end{equation}
where $\theta(\ )$ is the Heaviside function. Then, from 
(\ref{sc}) and (\ref{s0def}), we can express $\tilde S_0(\beta)$ 
in terms of $C(\beta)$, using the condition 
\begin{equation}
\e^{\tilde S_0(0)+o(N)}= |\Omega|^{N}.
\end{equation}
Combining this result with (\ref{zest}) and 
(\ref{s0def}), we rewrite the second and third 
terms of (\ref{hks:def4}) as 
\begin{equation}
-f(N) \equiv \log \frac{|\Omega|}{(\pi\e)^{d/2}}
-\frac{1}{N}\left[ \int_0^\infty d\beta 
\frac {C(\beta)-C_\infty \theta(\beta-1)}{\beta} \right],
\label{gamma:def}
\end{equation}
where we have used the relation  
$
\lim_{\beta \to \infty} \beta E_*(\beta)=C_\infty=Nd/2.
$
Using this quantity, the KS entropy given by (\ref{hks:def4}) 
can be expressed as
\begin{equation}
\hks= \lim_{N \to \infty} [ 
\lim_{\beta \to \infty} \bra \lambda \ket_{\beta}-f(N)
].
\label{hks:def6}
\end{equation}
Therefore, the KS entropy is obtained from $\bra \lambda \ket_{\beta}$
and $C(\beta)$, which can be computed using 
the canonical ensemble (\ref{can}) 
with the Hamiltonian (\ref{hamil:def}).

As one method to obtain the canonical ensemble, 
we consider the Langevin equation
\begin{equation}
\dot \Gamma_n(t)=-\pder{H([\Gamma])}{\Gamma_n}+\xi_n,
\label{lan}
\end{equation}
where  $\xi_n$ represents Gaussian white noise satisfying
\begin{equation}
\bra \xi_n(t)\xi_m(t')\ket=2 \beta^{-1} \delta(t-t')\delta_{nm}.
\end{equation}
The steady distribution of arrays $[\Gamma]$ in this system 
is given by  
the canonical ensemble (\ref{can}) with the Hamiltonian (\ref{hamil:def}).
However, because there are $O(\e^{N})$ ground states of $H$, 
it is difficult to obtain accurately the canonical ensemble (\ref{can}) by  
direct numerical integration of (\ref{lan}) when $\beta$ is large.  
For this reason, we employ an exchange Monte Carlo method \cite{exmc} 
in order to improve the accuracy of the numerical computations. Below 
we describe this method. 

We first prepare  $L$ copies of the Langevin system. 
Each system is  subject  to 
noise with a time-dependent inverse temperature $\beta^{(i)}(t)$ 
that takes values in the set $\{\beta^{(1)}_0, \cdots, \beta^{(L)}_0\}$.
We have the initial values $\beta^{(i)}(0)=\beta^{(i)}_0$, 
and the time evolution of each 
$\beta^{(i)}(t)$
is described by the following rule: At time $k \tau$ with $k=1,2,\cdots$,  
two inverse 
temperatures $\beta^{(j)}$ and $\beta^{(j')}$, given by 
$\beta^{(j)}=\beta^{(i)}_0$ and $\beta^{(j')}=\beta^{(i+1)}_0$ for some 
$i$ satisfying  
$i=(1-(-1)^k)/2+2i'-1$, with 
$i'=1,2,\cdots$, are exchanged with  probability 
\begin{equation}
\frac{1}{1+\e^{(\beta^{(i+1)}_0-\beta^{(i)}_0)
(H([\Gamma]^{(i)})-H([\Gamma]^{(i+1)}))}},
\end{equation}
where $[\Gamma]^{(i)}$ is the configuration of the system subject  
to the noise with the inverse temperature $\beta_0^{(i)}$. It can be 
shown that the steady distribution of $([\Gamma]^{(1)},\cdots,
[\Gamma]^{(L)})$ satisfies 
\begin{equation}
P_{\rm c}([\Gamma]^{(1)},\cdots,[\Gamma]^{(L)})=\frac{1}{\tilde Z}
\e^{-\sum_{i=1}^L \beta^{(i)}_0 H([\Gamma]^{(i)})},
\end{equation}
where $\tilde Z$ is a normalization constant.  
It has been demonstrated that the exchange Monte Carlo method 
is useful for calculating thermodynamic quantities  
in systems with many metastable states \cite{glass}, and therefore 
we expect that this method will be effective  in application to 
the present problem. 

%%%%%%%%%%%%%%%%%%%%%%%%%%%%%%%%%%
% example                        %
%%%%%%%%%%%%%%%%%%%%%%%%%%%%%%%%%%

\begin{figure}
\twofigures[width=6cm]{hc2.eps}{lyap2.eps}
\caption{Heat capacity  as a function of $\log\beta$. The 
circles, squares, triangles, stars 
and pluses represent the results for the cases $N=4, 8, 12, 16$
and $20$.}
\label{fig1}
\caption{$\bra\lambda\ket_\beta$ as a function of $\log\beta$. 
Here, $N=20$.}
\label{fig2}
\end{figure}

%\begin{figure}
%\begin{center}
%\includegraphics[width=6cm]{hc2.eps}
%\end{center}
%\caption{Heat capacity  as a function of $\log \beta$. 
%The red, green, blue, pink 
%and aqua symbols represent the results for the cases $N=4, 8, 12, 16$
%and $20$.} 
%\label{fig1}
%\end{figure}

%\begin{figure}
%\begin{center}
%\includegraphics[width=6cm]{lyap2.eps}
%\end{center}
%\caption{$\bra \lambda \ket_\beta$ as a function of $\log\beta$ 
%where $N=20$.} 
%\label{fig2}
%\end{figure}

\section{Example:}

As a simple demonstration of the computational  method described above,
we study a map $F$ given by
\begin{eqnarray}
x_{n+1} &=& x_n+z_n, \nonumber   \\
z_{n+1} &=& z_n+\frac{4}{g}\cos(x_n+z_n),  
\end{eqnarray}
which is defined in the region 
$\Omega=\{(x,z)|0 \le x \le 2\pi, 0 \le x +z \le 2\pi \} 
%\subset \mbox{\boldmath$R$^2}$.    
\subset {\bf R}^2$. 
This map is obtained as an approximation of a two-dimensional
map that describes a bouncing ball on a vibrating table \cite{chaos2}.
It also becomes the so-called standard map under a variable transformation.
We wish to compute the KS entropy of the repeller of $\Omega$,  
which is defined as the set $\{\Gamma | F^{n}(\Gamma) \in \Omega  
{\rm\hspace{2mm} for\hspace{2mm} all}\hspace{2mm} n\in 
{\bf Z} \}$.   

Specifically, we investigate  the case $g=0.3$. 
In the numerical simulations, 
we set $\tau=10^{-3}$ and $\beta^{(i)}_0=1.4^{-(i-8)}$, where 
 $1 \le i \le L=30$,
with a time step $\Delta t=10^{-5}$ for the integration of 
(\ref{lan}).  The heat capacity, $C(\beta)/N$, is plotted for
$N=4,8,12,16$ and $20$ in Fig. \ref{fig1}. From the 
value of the heat capacity thus obtained numerically,  
we derive the estimation 
 $\lim_{N \to \infty} f(N)=1.8(3)$. 
Furthermore,  $\bra \lambda \ket_\beta$ as a function of $\beta$
is displayed in Fig. \ref{fig2}.  Using this result, we derive 
$\bra \lambda \ket_\infty=2.5(0)$. Thus, from (\ref{gamma:def}), we 
obtain $\hks=0.6(7)$.

Let us check the validity of our computational result. 
First, because the escape rate formula holds for hyperbolic  
maps \cite{KT}, the value $\lim_{N \to \infty} f(N)$ should be 
equal to the escape rate, $\gamma$. Directly computing  
the escape rate, we obtain $\gamma=1.8(3)$. Thus, our 
computational result is consistent with
the escape rate formula. Second, in this model, we can directly 
evaluate the right-hand side of (\ref{hks:def}) by using the binary 
symbols $0$ and  $1$ for $x <\pi$  and $x \ge \pi$, respectively.
Collecting  many segments of trajectories on the repeller over the 
time intervals  $-\tilde N \le n \le \tilde N$,  with 
for $\tilde N=1,2$ and $3$, we obtain $\hks=0.6(8)$.
This value is also consistent with that found using  our method.

%%%%%%%%%%%%%%%%%%%%%%%%%%%%%%%%%%
% discussion                     %
%%%%%%%%%%%%%%%%%%%%%%%%%%%%%%%%%%

\section{Discussion:}

Let us finally comment on topics related to the basic formula 
(\ref{hks:def4}). 
% thermodynamic formalism
First, the statistical mechanical formulation used in the present paper 
is not related to the so-called thermodynamic formalism of chaos \cite{BS}, 
in which, instead of (\ref{pk:def}), the probability
\begin{equation}
p_k=\frac{1}{Z_d(\tilde \beta,N)}\e^{-N\tilde \beta\lambda([\Gamma^k])}
\end{equation}
is assumed for each periodic orbit.  
In a manner similar to that used here, $\Zd(\tilde \beta,N)$ can 
be obtained with a  statistical mechanical formulation, 
but the Hamiltonian
for cases in which $\tilde \beta\not=1$ includes the Jacobian 
$J([\Gamma])$, and this makes it difficult to carry out the 
computation. 

% Aubry 

Second, it is well known that the chaotic trajectories of the 
standard map 
correspond to the stationary configurations of  the Frenkel-Kontorowa 
model \cite{Aubry}. However, such a correspondence can be found only 
for a certain class of maps. Also, even for the standard map, 
it seems difficult to develop a method for computing the KS entropy 
by using this correspondence, because  periodic orbits correspond 
to local minima and maxima, not ground states of the Frenkel-Kontorowa 
Hamiltonian. Nevertheless, it is worth noting that this correspondence
has led to a mathematical study to evaluate the lower bound of
the topological entropy \cite{Knill}. 

% non-hyperbolic point 

Finally, let us recall  that the expression (\ref{hks:def4}) is valid 
in the case that none of the  eigenvalues $\Lambda_m$ ($1 \le m\le d$)  
of the  periodic orbits is equal to $1$.  
However, even if there is a non-hyperbolic 
periodic orbit whose eigenvalue $\Lambda_m$ is unity, 
the right-hand side of the expression (\ref{hks:def4}) can be
computed.  It might be an interesting problem to investigate 
the expression (\ref{hks:def4})  for non-hyperbolic systems.

% conclusion 

In conclusion, 
we  derived  the  formulae (\ref{hks:def4})
and (\ref{hks:def6}) for the  KS entropy. We then applied  these 
formulae to compute the KS entropy 
for the case of the 
repeller of a simple map by using the corresponding  
Langevin equation together  
with the exchange Monte Carlo method. We hope that these formulae 
can be similarly applied to many  problems. 

\acknowledgments

The authors  acknowledge K. Hukushima for teaching them some techniques
of the exchange Monte Carlo method. They also thank Y. Ishii for
informing them of Ref. \cite{Knill}. 
This work was supported by a grant from the Ministry of 
Education, Science, Sports and Culture of Japan (No. 16540337) and 
a Research Fellowship for Young Scientists from 
the Japan Society for the Promotion of Science (No. 1711222).

\end{document}